\documentclass[%
 aip,
 jcp,
 amsmath,amssymb,
 reprint,
 normalem,%
]{revtex4-1}

\usepackage{graphicx}
\usepackage{dcolumn}
\usepackage{bm}
\usepackage{dsfont}
\usepackage{xcolor}
\usepackage{ulem}
\usepackage[utf8]{inputenc}
\usepackage[T1]{fontenc}
\usepackage{mathptmx}
\usepackage{mathtools}
\usepackage{etoolbox}
\usepackage{tabularx}
\usepackage{placeins}
\usepackage{xspace}
\usepackage{booktabs}
\usepackage{multirow}
\usepackage[hidelinks]{hyperref}

\makeatletter
\def\@email#1#2{%
 \endgroup
 \patchcmd{\titleblock@produce}
  {\frontmatter@RRAPformat}
  {\frontmatter@RRAPformat{\produce@RRAP{*#1\href{mailto:#2}{#2}}}\frontmatter@RRAPformat}
  {}{}
}%
\makeatother
\begin{document}

\title[]{FMO-LC-TDDFTB method for excited states of large molecular assemblies in the strong light-matter coupling regime}

\author{Richard Einsele}
  \altaffiliation{The authors R.E. and L.N.P. contributed equally to this paper.}
\author{Luca Nils Philipp}
  \altaffiliation{The authors R.E. and L.N.P. contributed equally to this paper.}
\author{Roland Mitri\'c}
\email{roland.mitric@uni-wuerzburg.de}
\affiliation{Institut für Physikalische und Theoretische Chemie, Julius-Maximilians-Universität Würzburg, Emil-Fischer-Strasse 42, 97074 Würzburg, Germany}
\date{\today}

\begin{abstract}
We present a new methodology to calculate the strong light-matter coupling between photonic modes in microcavities and large molecular aggregates that consist of hundreds of molecular fragments. To this end, we combine our fragment molecular orbital long-range corrected time-dependent density functional tight-binding methodology with a generalized Tavis-Cummings Hamiltonian. We employ an excitonic Hamiltonian, which is built from a quasi-diabatic basis that is constructed from locally excited and charge-transfer states of all molecular fragments. In order to calculate polaritonic states, we extend our quasi-diabatic basis to include photonic states of a microcavity and derive and implement the couplings between the locally excited states and the cavity states and built a Tavis-Cummings Hamiltonian that incorporates the intermolecular excitonic couplings. Subsequently, we demonstrate the capability of our methodology by simulating the influence of the electric field polarization on the polaritonic spectra for a tetracene aggregate of 125 monomers. Furthermore, we investigate the dependence of the splitting of the upper and lower polaritonic branches on the system size by comparing the spectra of three different tetracene clusters. In addition, we investigate the polariton dispersion of a large tetracene aggregate for electric field polarizations in the x, y and z direction. Our new methodology can facilitate the future study of exciton dynamics in complex molecular systems, which consist of up to hundreds of molecules, that are influenced by strong light-matter coupling to microcavities.
\end{abstract}

\maketitle

\section{Introduction}
Molecular polaritons are hybrid light-matter states which arise by the strong-coupling between molecular excited states and quantized modes of confined electromagnetic fields. Recent observations of novel physical and chemical dynamics induced by polariton effects showed that strong-coupling is an effective way to manipulate molecular properties without applying synthetic modifications, which sparked great interest in the field and lead to numerous theoretical and experimental studies investigating strong light-matter interactions.\cite{nagarajan_chemistry_2021, hutchison_modifying_2012, thomas_tilting_2019, garcia-vidal_manipulating_2021,damari_strong_2019,vergauwe_modification_2019,berghuis_controlling_2022,rozenman_long-range_2018,akulov_long-distance_2018,flick_ab_2018,csehi_ultrafast_2019,mandal_investigating_2019,golombek_collective_2020,qiu_molecular_2021,ye_direct_2021,balasubrahmaniyam_coupling_2021,gudem_cavity-modified_2023,godsi_exploring_2023,pajunpaa_polariton-assisted_2024,wang_situ_2021}

While a large area of this research is focused on molecular polaritons, which are formed by the strong-coupling of molecular vibrations\cite{damari_strong_2019,lather_cavity_2019,thomas_tilting_2019,vergauwe_modification_2019,pang_role_2020,thomas_ground_2020,fukushima_vibrational_2021,joseph_supramolecular_2021,li_cavity_2021,nagarajan_chemistry_2021,fukushima_inherent_2022,tichauer_identifying_2022}, this work will concentrate on the coupling of electronic excitations to confined electromagnetic field modes.
This area of the polaritonic research investigates a multitude of interesting phenomena, among which are control of chemical reaction rates\cite{hutchison_modifying_2012, schwartz_reversible_2011} and enhanced charge and energy transfer in organic semiconductors\cite{orgiu_conductivity_2015, hagenmuller_cavity-enhanced_2017, lerario_high-speed_2017, hou_ultralong-range_2020, feist_extraordinary_2015, schachenmayer_cavity-enhanced_2015}.
However, many of the experimental observations are not well understood\cite{hutchison_modifying_2012, schwartz_polariton_2013, georgiou_generation_2018, liu_role_2020}, which calls for accurate theoretical models to gain insight into the underlying microscopic processes. 

The theoretical description of polaritonic systems necessitates an accurate treatment of both the photonic and the molecular degrees of freedom.
Subsequently, the incorporation of light-matter interactions into theoretical approaches is required. To this end, the description of the electric field within the framework of quantum electrodynamics (QED) is combined with quantum chemical methods to facilitate the simulation of molecular/realistic systems.
Usually, this is done in either of two ways. Within the self-consistent QED (scQED) approach the polaritonic states are directly obtained from a single self-consistent field procedure, which includes photonic as well as electronic degrees of freedom.\cite{schafer_ab_2018,flick_lightmatter_2019,sidler_perspective_2022,ruggenthaler_understanding_2023,mandal_theoretical_2023}
In the second approach, adiabatic electronic states are initially calculated and subsequently used as a basis for a QED Hamiltonian to obtain polaritonic states. This methodology will be referred to as the parameterized QED (pQED) approach.

While QED Hamiltonians are easily combined with most quantum chemical methods within the pQED scheme, the implementation within the scQED scheme is more complicated. Nevertheless, most commonly used quantum chemical methods have been extended to account for strong-coupling within the scQED scheme for ground as well as excited states , e.g., QED HF\cite{haugland_coupled_2020}, QED (EOM-)CC\cite{haugland_coupled_2020, pavosevic_polaritonic_2021}, and QED (TD-)DFT\cite{tokatly_time-dependent_2013, ruggenthaler_quantum-electrodynamical_2014,flick_kohnsham_2015}. While these methods should, in general, provide the most accurate description of polaritonic ground and excited states, they are often limited to the single molecule and single cavity mode cases as well as the long-wavelength approximation due to computational costs. However, most actual experiments in the molecular strong-coupling regime are carried out by coupling large molecular assembles to many modes of micro-cavities or surface plasmon polaritons (SPP).\cite{berghuis_controlling_2022, schwartz_reversible_2011, hutchison_modifying_2012, orgiu_conductivity_2015}

Due to the lack of methods, which can accurately describe such large polaritonic systems, many theoretical studies in the collective strong-coupling regime approximate large molecular aggregates as isolated and non-interacting molecules.
Then, within a pQED scheme, the adiabatic electronic states of every single molecule in the aggregate are calculated and afterwards, all adiabatic electronic states of the molecules are used to build a (generalized) Tavis-Cummings Hamiltonian.\cite{luk_multiscale_2017, sokolovskii_photochemical_2024, berghuis_controlling_2022, sokolovskii_non-hermitian_2024} 

The Tavis-Cummings (TC) model is the most commonly used QED Hamiltonian to describe the interaction between an ensemble of identical, non-interacting two-level emitters and a single mode of an electromagnetic field.\cite{tavis_exact_1968, tavis_approximate_1969} It is straightforward to generalize the model to account for the multi-level nature of molecules, multiple modes of electromagnetic fields and to include the spatial variation of the electromagnetic field. The TC Hamiltonian has been employed in various studies to investigate the influence of strong light-matter interactions on molecular systems\cite{tichauer_identifying_2022,delpo_polariton_2020,groenhof_coherent_2018,hulkko_effect_2021,tichauer_multi-scale_2021,tichauer_tuning_2023,sokolovskii_multi-scale_2023}.

However, compared to other QED Hamiltonians, it neglects the dipole self-energy, counter rotating-wave terms and permanent dipole moments of the molecules. Thus, it is expected that the TC Hamiltonian breaks down for ultrastrong coupling strengths. Furthermore, intermolecular interactions play a crucial role in energy and charge transport properties of excitonic materials and therefore, have to be included for an accurate description of these processes in polaritonic systems.\cite{maly_wavelike_2020}

Thus, the combination between an excitonic Hamiltonian for the description of the intermolecular excitonic couplings\cite{nottoli_role_2018,green_excitonic_2021,canola_addressing_2021,einsele_long-range_2023} and the TC model would enable a more accurate invesitgation of polaritonic systems. 
Our recently developed fragment molecular orbital long-range corrected time-dependent density functional tight-binding methodology (FMO-LC-TDDFTB)\cite{einsele_long-range_2023}, which facilitates the efficient calculation of the electronically excited states of large molecular assemblies consisting of hundreds of molecules, can be easily combined with a TC model to simulate the excited-state spectra of cavity coupled systems. The methodology is based on the fragment molecular orbital (FMO)\cite{kitaura_fragment_1999,nakano_fragment_2000,nakano_fragment_2002,fedorov_multilayer_2005,nishimoto_density-functional_2014} theory, which represents a fragmentation approach, where the complete molecular system is partitioned into numerous fragment. The properties of the full system are then obtained by combining the results of the calculations of the isolated fragments and their interactions between each other.

In our method, we employ a quasi-diabatic basis, which consists of locally excited (LE) and charge-transfer (CT) states that are calculated with the LC-TD-DFTB methodology for monomer and pair fragments. 
The excitonic Hamiltonian is constructed from the energies of the basis states and the excitonic couplings that are calculated in the tight-binding formalism. The excited-state energies of the complete system are then obtained by diagonalizing the Hamiltonian.

In this work, we adapt our FMO-LC-TDDFTB method to account for strong-coupling to photons in confined electromagnetic environment by combining it with a generalized TC Hamiltonian. To this end, we expand the quasi-diabatic basis of LE and CT states with the states of the electromagnetic field modes. Couplings between the quasi-diabatic fragment excited states and photonic states are calculated from the transition dipole moments of the respective electronic transitions within the approximations of the TC Hamiltonian and subsequently, the excited states of the full system are obtained by diagonalization of the resulting polaritonic Hamiltonian matrix. By applying our method to large clusters of tetracene molecules strongly coupled to an electromagnetic field, we show that we can gain insights lying beyond the typically employed coupled oscillator models\cite{novotny_strong_2010}. 

The present work is structured as follows: in section \ref{sec:methodology}, the theoretical framework of our method is defined. The computational details for the calculations that are perfomed in this paper are given in section \ref{sec:comp_details}. The results of the simulations of the strong-light matter coupling in the tetracene aggregates are discussed in section \ref{sec:results}. 
Finally, in section \ref{sec:conclusions}, conclusions and an outlook are presented.
\newcommand\numberthis{\addtocounter{equation}{1}\tag{\theequation}}
\section{Methodology}\label{sec:methodology}
In this chapter, we will derive the working equations for the calculation of the strong light-matter coupling between a cavity and molecular aggregates. At first, the FMO-LC-DFTB formalism for the electronic ground state \cite{nishimoto_density-functional_2014} is presented. Thereafter, we give a summary of the theoretical methodology of the FMO-LC-TDDFTB method for the calculation of the electronically excited-states of large molecular systems.

Subsequently, the TC model for the description of strong-coupling between an ensemble of two-level systems and multiple electromagnetic field modes is introduced. This is followed by the derivation of excited polaritonic states of molecular aggregates.

In this work, we employ the following notation convention: uppercase letters A, B and C denote atoms, whereas molecular fragments are delineated by uppercase letters I through L without indices. Matrix elements are designated by uppercase letters with indices, while matrices are depicted by bold uppercase letters. Indices denoting molecular orbitals are expressed using lowercase letters, with Greek letters reserved for atomic orbital indices. Throughout the formulation of the FMO-LC-TDDFTB methodology, atomic units are employed.
All of the methodology has been implemented in our software DIALECT, which is available on github.\cite{noauthor_dialect_2024}

\subsection{FMO-LC-DFTB}\label{sec:FMO-LC-DFTB}
Nishimoto, Fedorov and Irle introduced the combination of the fragment molecular orbital method with the density functional tight-binding method to simulate extensive molecular systems\cite{nishimoto_density-functional_2014}. Vuong \textit{et al} adapted the methodology to LC-DFTB\cite{vuong_fragment_2019}, which we employ in our methodology to calculate ground-state properties. Thus, in this section, we will only briefly summarize the most important equations of the method and refer to the aforementioned works\cite{nishimoto_density-functional_2014,vuong_fragment_2019} for a detailed formulation.
The total ground-state energy of in the FMO-LC-DFTB methodology is defined as
\begin{align}\label{eq:fmo_ground_state}
\begin{split}
    E=&\sum_I^N E_I+ \sum_I^N \sum_{J > I}^N\left(E_{I J}-E_I-E_J\right)
    + \sum_I^N \sum_{J > I}^N \Delta E_{I J}^{\mathrm{em}},
\end{split}
\end{align}
where
\begin{align}
\begin{split}
    E_X =& \sum_{\mu \nu} P_{\mu v} H_{\mu v}^0 
 +\frac{1}{2} \sum_{\mu, \sigma, \lambda, v} \Delta P_{\mu \sigma} \Delta P_{\lambda \nu}(\mu \sigma \mid \lambda v)  \\
& -\frac{1}{4} \sum_{\mu, \sigma, \lambda, v} \Delta P_{\mu \sigma} \Delta P_{\lambda v}(\mu \lambda \mid \sigma v)_{\mathrm{lr}} 
 +\sum_{A, B} V_{A B}^{\mathrm{rep}}\left(R_{A B}\right)
\end{split}
\end{align}
is the internal LC-DFTB energy of a single monomer or pair fragment of the complete system. Here, $E_{I}$ ($E_{J}$) is the energy of a monomer fragment and $E_{IJ}$ the energy of a pair fragment. Applying the tight-binding formalism to the two-electron integrals in the Coulomb and exchange energy contributions results in the following expressions 
\begin{align*}\label{eq:2e_approx_ao}
(\mu \lambda \mid \sigma \nu) &=\iint \phi_{\mu}(r_1) \phi_{\lambda}(r_1) \frac{1}{r_{12}} \phi_{\sigma}(r_2) \phi_{\nu}(r_2) \mathrm{d} r_1 \mathrm{d} r_2 \\
& \approx \sum_{A, B} \gamma_{A B} q_{A}^{\mu \lambda} q_{B}^{\sigma \nu}, \numberthis
\end{align*}
and 
\begin{align*}\label{eq:2e_approx_ao_lr}
(\mu \lambda \mid \sigma \nu)_{lr} &=\iint \phi_{\mu}(r_1) \phi_{\lambda}(r_1) \frac{\mathrm{erf}(\frac{r_{12}}{R_{lr}})}{r_{12}} \phi_{\sigma}(r_2) \phi_{\nu}(r_2) \mathrm{d}r_1 \mathrm{d}r_2 \\
& \approx \sum_{A, B} \gamma_{A B}^{lr} q_{A}^{\mu \lambda} q_{B}^{\sigma \nu}. \numberthis
\end{align*}
The $\bm{\gamma}$-matrices, which represent charge fluctuation interactions, are calculated by employing spherical Gaussian functions
\begin{equation}
\gamma_{A B}=\frac{\operatorname{erf}\left(C_{A B} R\right)}{R},
\label{eq:gamma}
\end{equation}
where $R$ represents the distance between the atoms $A$ and $B$. $C_{A B}=(2\left(\sigma_A^2+\sigma_B^2\right))^{-\frac{1}{2}}$ depends on the widths $\sigma_A$ and $\sigma_B$ of the charge clouds on the two atoms. Atom-specific Hubbard parameters $U_A$ as $\sigma_A= (\sqrt{\pi} U_A)^{\frac{1}{2}}$ determine the widths.
The long-range $\gamma$-matrix is given by
\begin{equation}
\gamma_{A B}^{\operatorname{lr}}=\frac{\operatorname{erf}\left(C_{A B}^{\mathrm{lr}} R\right)}{R},
\end{equation}
where
\begin{equation}
C_{A B}^{\mathrm{lr}}=\frac{1}{\sqrt{2\left(\sigma_A^2+\sigma_B^2+\frac{1}{2} R_{\mathrm{lr}}^2\right)}}
\end{equation}
depends on the range-separation parameter $R_{\mathrm{lr}}$. The transition charges $q_{A}^{\mu \lambda}$ of atom A are
\begin{equation}
q_{A}^{\mu \lambda}=\frac{1}{2}(\delta(\mu \in A)+\delta(\lambda \in A)) S_{\mu \nu}.
\label{eq:q_between_aos}
\end{equation}
The difference between the embedding energy of the pair and the monomers is given by
\begin{align}\label{eq:embedding_energy}
\begin{split}
    \Delta E_{I J}^{\mathrm{em}}&=E_{I J}^{\mathrm{em}}-E_I^{\mathrm{em}}-E_J^{\mathrm{em}}\\
    &=\sum_{A \in I J} \sum_{K \neq I, J}^N \sum_{C \in K} \gamma_{A C} \Delta \Delta q_A^{I J} \Delta q_C^K.
\end{split}
\end{align}
Here, the expression
\begin{align}
\Delta q_A^{K} & =q_A^{K}-q_A^{0,K} \\
& =\sum_{\mu \in A} \sum_v\left[P_{\mu v}^{K} S_{\mu v}^{K}-P_{\mu v}^{0,K} S_{\mu v}^{K}\right]
\end{align}
describes the Mulliken charge difference of atom A, $\Delta \Delta q_A^{I J}$ denotes the difference between the Mulliken charge of a pair fragment and its containing monomers\cite{nishimoto_density-functional_2014,einsele_long-range_2023}. 

The electrostatic-dimer (ES-DIM)\cite{nakano_fragment_2002} approximation is employed for far-separated fragment pairs pairs, whose orbital overlap will be zero. The energy of ES-DIM pairs is defined as
\begin{equation}\label{eq:ESDIM_energy}
    E_{I J}=E_I+E_J+\sum_{A \in I} \sum_{B \in J} \gamma_{A B} \Delta q_A^I \Delta q_B^J.
\end{equation}
The ground-state Hamiltonian 
\begin{align}
\begin{split}
    H_{\mu \nu}^X&=H_{\mu \nu}^{0}+\frac{1}{2} S_{\mu \nu} \sum_{C}\left(\gamma_{A C}+\gamma_{B C}\right) \Delta q_{C} \\
&+V_{\mu \nu}^X -\frac{1}{8} \sum_{\alpha \beta} \Delta P_{\alpha \beta} S_{\mu \alpha} S_{\beta \nu} \\
& \times \left(\gamma_{\mu \beta}^{\mathrm{lr}}+\gamma_{\mu \nu}^{\mathrm{lr}}+\gamma_{\alpha \beta}^{\mathrm{lr}}+\gamma_{\alpha \nu}^{\mathrm{lr}}\right),
\end{split}
\end{align}
of a singular fragment in the framework of the FMO-DFTB method incorporates the Coulomb interaction to all other fragments of the system, which is given by 
\begin{equation}
    V_{\mu v}^X=\frac{1}{2} S_{\mu \nu}^X \sum_{K \neq X}^N \sum_{C \in K}^{N_K}\left(\gamma_{A C}+\gamma_{B C}\right) \Delta q_C^K.
\end{equation}
Therefore, the calculation of ground-state properties in the FMO-DFTB method necessitates concurrent self-consistent charge (SCC) calculations of all monomers, where the potential $V_{\mu v}^X$ is updated in each successive iteration. A more detailed depiction of the implementation of the monomer SCC iterations is presented in earlier work of our group.\cite{einsele_long-range_2023}

\subsection{FMO-LC-TDDFTB}\label{sec:FMO-LC-TDDFTB}
In our previous work, we formulated the FMO-LC-TDDFTB methodology to calculate the electronically excited states of large molecular systems\cite{einsele_long-range_2023}. Therefore, we will only provide a summary of the most important equations of the methodology in this section. For details, we refer to our previous publication.\cite{einsele_long-range_2023}

The excited state wavefunction in the framework of the FMO-LC-TDDFTB method can be expressed as a linear combination of locally excited states on single fragments and charge-transfer states between two fragments, yielding the following expression for the total electronic excited state wavefunction  
\begin{equation}\label{eq:exc_state_wf}
\left|\Psi\right\rangle=\sum_{I}^N \sum_{m}^{N_{\mathrm{LE}}} c_I^m \left|\mathrm{LE}_I^m \right\rangle+ \sum_{I}^{N} \sum_{J \neq I}^{N} \sum_{m}^{N_{\mathrm{CT}}} c_{I\rightarrow J}^m \left| \mathrm{CT}_{I \to J}^{m}\right\rangle.
\end{equation}
Here, the coefficients of the quasi-diabatic basis LE and CT basis states are obtained by solving the eigenvalue equation $\mathbf{Hc} = \mathbf{Ec}$.
The LE states are calculated as singlet excited states of monomer fragments and can be defined as
\begin{equation}
    \left|\mathrm{LE}_{\mathrm{I}}^m\right\rangle = \sum_{i \in I}\sum_{a \in I}  X_{ia}^{m(I)} | \Phi_{I}^{i\to a}\rangle,
    \label{eq:LE_states}
\end{equation}
where $X_{ia}^{m(I)}$ is the transition density matrix of the $m$-th excited state of the $I$-th fragment in the MO basis of the respective monomer and $|\Phi_{I}^{i\to a}\rangle$ is the excited-state determinant of the excitation from the occupied orbital i to the virtual orbital a on fragment $I$.

The charge-transfer state of a fragment pair, where the excitation occurs from  monomer $I$ to monomer $J$, is defined as
\begin{align}
    \left|\mathrm{CT}_{I \to J}^{m}\right\rangle = \sum_{i \in I}\sum_{a \in J} X_{ia}^{m(I \to J)} | \Phi_{I \to J}^{i\to a} \rangle.
    \label{eq:CT_states}
\end{align}
Thus, the electron transfer between the monomers is limited to the occupied orbitals of monomer $I$ and the virtual orbitals of monomer $J$. 

The electronically excited-state energies of the quasi-diabatic basis states are obtained from linear-response calculations in the framework of the LC-DFTB method.\cite{humeniuk_long-range_2015} The excited states can be obtained by solving the so called Casida eigenvalue equation
\begin{equation}
    \left(\begin{array}{ll}
\mathbf{A} & \mathbf{B} \\
\mathbf{B} & \mathbf{A}
\end{array}\right)\left(\begin{array}{l}
\mathbf{X} \\
\mathbf{Y}
\end{array}
\right)=\bm{\Omega}\left(\begin{array}{cc}
\mathbf{1} & 0 \\
0 & -\mathbf{1}
\end{array}\right)\left(\begin{array}{l}
\mathbf{X} \\
\mathbf{Y}
\end{array}\right),\label{eq:Casida}
\end{equation}
where the matrices $\mathbf{A}$ and $\mathbf{B}$ are defined as
\begin{align}
\begin{split}
    A_{i a, j b} &=\delta_{i j} \delta_{a b}\left(\epsilon_a-\epsilon_i\right)+2 \sum_{A}\sum_{B} q^{ia}_A \gamma_{AB} q_B^{jb} \\
    &- \sum_{A} \sum_{B} q^{ij}_A \gamma_{AB}^{\mathrm{lr}} q^{ab}_B
\end{split}
\\
    B_{i a, j b} &= 2 \sum_{A}\sum_{B} q^{ia}_A \gamma_{AB} q_B^{jb} - \sum_{A} \sum_{B} q^{ib}_A \gamma_{AB}^{\mathrm{lr}} q^{aj}_B.
\end{align}
Here, the atomic transition charges between MOs are expressed as
\begin{equation}
q_{A}^{i j}=\frac{1}{2} \sum_{\mu \in A} \sum_{\nu}\left(C_{\mu i} C_{\nu j}+C_{\nu i}C_{\mu j}\right) S_{\mu \nu}.
\label{eq:q_between_mos}
\end{equation}
However, in our approach, we employ the the Tamm-Dancoff (TDA) approximation in the excited-state calculations, and thus, Eq. \ref{eq:Casida} is reduced to the Hermitian eigenvalue problem
\begin{equation}
    \mathbf{A} \mathbf{X} = \bm{\Omega} \mathbf{X},
\end{equation}
which can be solved by employing the Davidson algorithm to calculate the few lowest eigenvalues \cite{davidson_iterative_1975}.

The excited states of the complete system in the framework of the FMO-LC-TDDFTB method are calculated by diagonalizing the complete excited-state Hamiltonian, which is constructed from the quasi-diabatic LE and CT basis states of all fragments. Whereas the diagonal elements of the Hamiltonian are represented by the excited-state energies of the respective basis states, the off-diagonal matrix elements are obtained from the calculation of the excitonic couplings between the LE and CT states.
As formulated in Refs. \citep{fujita_development_2018,einsele_long-range_2023}, the off-diagonal matrix elements are partitioned in one-electron 
\begin{equation}\label{One_electron_hamiltonian}
    \langle \Phi_{I \to J}^{i\to a}|\mathrm{H}_{1e}|\Phi_{K \to L}^{j \to b} \rangle = \delta_{I K} \delta_{i j} H_{a b}^{\prime}-\delta_{J L } \delta_{a b} H_{i j}^{\prime}
\end{equation}
and two-electron contributions
\begin{align}\label{two_electron_hamiltonian}
\begin{split}
    \langle \Phi_{I \to J}^{i\to a} | \mathrm{H}_{2 e} |\Phi_{K \to L}^{j \to b} \rangle &=2\left(i^{(I)} a^{(J)} \mid j^{(K)} b^{(L)}\right)\\ 
    &-\left(i^{(I)} j^{(K)} \mid a^{(J)} b^{(L)}\right).
\end{split}
\end{align}
As the one-electron interaction vanishes for the LE-LE coupling according to the Slater-Condon rules, the coupling between the LE states is limited to the two-electron interactions, and thus, yields the following expression in the tight-binding formalism
\begin{align*}\label{le_le_coupling}
\left\langle \mathrm{LE}_{I}^{m}\right|\mathrm{H}& \left| \mathrm{LE}_{J }^{n}\right\rangle 
= 2 \sum_{A\in I} \sum_{B \in J} q_{\mathrm{tr}, A}^{m(I)} \gamma_{AB} q_{\mathrm{tr}, B}^{n(J)} \numberthis \\
-& \sum_{A\in IJ} \sum_{B \in IJ} 
\sum_{ia \in I} \sum_{jb \in J} X_{ia}^{m(I)} X_{jb}^{n(J)} q_{A}^{ij} \gamma_{AB}^{\mathrm{lr}} q_{B}^{ab},
\end{align*}
where
\begin{equation}
    q_{\mathrm{tr}, A}^{m(I)} = \sum_{ia} q_{A}^{ia} X_{ia}^{m(I)} 
\end{equation}
is the transition charge of the $m$-th excited state of the fragment $I$ on atom $A$.
While the first term of Eq. (\ref{le_le_coupling}) denotes the Coulomb interaction between the transition densities of fragments $I$ and $J$, the second terms represents the exchange interaction. In case of the ES-DIM approximation for far separated fragments, the exchange contribution to the LE-LE coupling is neglected.

The coupling between the LE state on fragment $I$ and the charge-transfer state of the excitation from fragment $J$ to $K$ is defined as
\begin{align}\label{le_ct_coupling}
\begin{split}
\left\langle\mathrm{\mathrm{LE}}_{I}^{m}\right|\mathrm{H}& \left| \mathrm{CT}_{J \to K}^{n}\right\rangle =  \delta_{I J} \sum_{ia \in I}\sum_{b \in K} X_{ia}^{m(I)} X_{ib}^{n(I \to K)} H_{a b}^{\prime} \\ -&\delta_{I K}
\sum_{ia \in I}\sum_{j \in J} X_{ia}^{m(I)}X_{ja}^{n(J \to I)} H_{i j}^{\prime} \\ 
+& 2\sum_{A \in I}\sum_{B \in JK} q_{\mathrm{tr}, A}^{m(I)} \gamma_{AB} q_{\mathrm{tr}, B}^{n(J \to K)} \\
- &\sum_{A \in IJ}\sum_{B \in IK
} \sum_{ia \in I}  \sum_{j \in J}\sum_{b \in K}X_{ia}^{m(I)} X_{jb}^{n(J \to K)} q_{A}^{ij} \gamma_{AB}^{\mathrm{lr}} q_{B}^{ab},  
\end{split}
\end{align}
where $H_{a b}^{\prime}$ are matrix elements of the orthogonalized ground-state Hamiltonian of the complete system, which is defined as
\begin{equation}
    \mathbf{H}^{\prime} = \mathbf{S}^{-\frac{1}{2}} \mathbf{H}^{LCMO} \mathbf{S}^{-\frac{1}{2}}.
    \label{eq:orthogonalized_hamiltonian}
\end{equation}
Here, $\mathbf{S}$ is the overlap matrix of the full system and the non-orthogonalized ground-state Hamiltonian is constructed from the Hamiltonian matrices of the monomer and pair fragments as
\begin{equation}
    \mathbf{H}^{LCMO} = \bigoplus_{I} \mathbf{H}^{I} + \bigoplus_{I>J}(\mathbf{H}^{IJ} - \mathbf{H}^{I} \oplus \mathbf{H}^{J}).
\end{equation}
The inverse of the square root of the overlap matrix is approximated according to the following expression
\begin{equation}
    \mathbf{S}^{-\frac{1}{2}} \approx \frac{3}{2} \mathbf{1} - \frac{1}{2} \mathbf{S}
\end{equation}
to decrease the computational demand of our methodology.\cite{einsele_long-range_2023}
In contrast to the LE-LE coupling, the one-electron contribution to the LE-CT coupling is not zero. However, if the LE and CT state are not situated on the same monomer and the ES-DIM approximation is applied in the case of far separated fragments, the contributions vanishes as the matrix elements of the orthogonalized ground-state Hamiltonian $\mathbf{H}^{'}$ (\textit{cf}. Eq. \ref{eq:orthogonalized_hamiltonian}) are zero for ES-DIM pair fragments. Additionally, if the ES-DIM approximation is employed, the only remaining contribution of the LE-CT coupling is the Coulomb interaction
\begin{equation}
    \left\langle\mathrm{\mathrm{LE}}_{I}^{m}\left|\mathrm{H}\right| \mathrm{CT}_{J \to K}^{n}\right\rangle = 2\sum_{A \in I}\sum_{B \in JK} q_{\mathrm{tr}, A}^{m(I)} \gamma_{AB} q_{\mathrm{tr}, B}^{n(J \to K)}.
\end{equation}

In the case of the couplings between CT states, the one-electron contribution is included in the excited-state energies of the respective basis states that are obtained from the LC-TDDFTB calculations. The remaining two-electron interaction of the couplings between two CT states is given by
\begin{align}\label{ct_ct_coupling}
\begin{split}
\left\langle \mathrm{CT}_{I \to J}^{m} \right|& \mathrm{H}\left| \mathrm{CT}_{K \to L}^{n}\right\rangle = 2 \sum_{A\in IJ} \sum_{B \in KL} q_{\mathrm{tr}, A}^{m(I \to J)} \gamma_{AB} q_{\mathrm{tr}, B}^{n(K \to L)} \\
&- \sum_{i \in I}\sum_{a \in J}\sum_{j \in K}\sum_{b \in L} \sum_{A \in IK} \sum_{B \in JL} X_{ia}^{m(I \to J)}\\
&\times X_{jb}^{n(K \to L)}q_{A}^{ij} \gamma_{AB}^{\mathrm{lr}} q_{B}^{ab}. 
\end{split}
\end{align}
If the fragments $I$ and $K$ or $J$ and $L$ are not in close spatial proximity, the ES-DIM approximation is applied and thus, the exchange contribution of the CT-CT coupling is neglected.

\subsection{Tavis-Cummings model}\label{sec:Tavis-Cummings}

Strong-light matter interactions between large ensembles of molecules and electromagnetic fields are usually treated within the (generalized) TC model. It describes the interaction between an ensemble of \textit{N} systems and multiple modes of an quantized electromagnetic field. The corresponding Hamiltonian is given by

\begin{align}
\begin{split}
    H_{TC} = \hbar\sum_{I}^N\omega_{\mathrm{mol},I}\sigma_I^+\sigma_I^- + \hbar \sum_\lambda\omega_{\mathrm{ph},\lambda}a^\dag_\lambda a_\lambda  \\ 
    + \hbar \sum_{I}^N\sum_\lambda g_{I,\lambda} \left(a_\lambda^\dag\sigma_I^- + \sigma_I^+a_\lambda\right),\label{TCH}
\end{split}
\end{align}
where $a_\lambda$ ($a_\lambda^\dag$) is the annihilation (creation) operator of the quantized electromagnetic field mode of frequency $\omega_{\mathrm{ph},\lambda}$ and $\sigma_I^-$ ($\sigma_I^+$) is the Pauli raising (lowering) operator of the \textit{I}-th system of frequency $\omega_{\mathrm{mol},I}$.

The light-matter coupling constant $g_{I,\lambda}$ between the \textit{I}-th system and the $\lambda$-th mode of the electromagnetic field is given by

\begin{equation}
    g_{I,\lambda} = \sqrt{\frac{\hbar\omega_{\mathrm{ph},\lambda}}{2\epsilon_0 V}} \Vec{\mu}_{eg}^{(I)}\cdot\Vec{\epsilon}_\lambda,\label{coupling}
\end{equation}
where $V$ is the quantized volume of the electromagnetic field, $\Vec{\epsilon}_\lambda$ is the polarization vector of the $\lambda$-th mode of the electromagnetic field, and $\Vec{\mu}_{eg}^{(I)}$ is the transition dipole moment of the ground to excited state transition of the \textit{I}-th system.

Compared to other commonly used QED Hamiltonians, like the Dicke \cite{hepp_superradiant_1973, dicke_coherence_1954} or Pauli-Fierz \cite{rokaj_lightmatter_2018, schafer_ab_2018} Hamiltonian, the TC Hamiltonian conserves the number of particles inside the system, i.e., the operator 
\begin{equation}
    N_p = \sum_\lambda a_\lambda^\dag a_\lambda + \sum_I^N \sigma_I^+\sigma_I^-
\end{equation}
commutes with the Hamiltonian. Thus, eigenstates of the TC Hamiltonian are also eigenstates of $N_p$ and basis states with different eigenvalues with respect to $N_p$ do not couple within the TC model. Therefore, the TC Hamiltonian has a block diagonal structure and the eigenstates with a certain number of excitations can be obtained by diagonalizing the corresponding block, which has finite dimensions. In the following, we will limit ourselves to the calculation of the single excitation eigenstates of the coupled system, since those are required for the interpretation of experiments under weak driving conditions, i.e., linear absorption measurements.

We will first discuss the solution of the Tavis-Cummings model with a single electromagnetic field mode and identical two-level systems, i.e., $\omega_{\mathrm{mol}} = \omega_{\mathrm{mol},I}$ and $g = g_{1,\lambda}$, to develop an understanding of the eigenstate structure and introduce the commonly used notation.
Conveniently, this model is solved in the direct product basis of the bare molecular states and the eigenstates of the cavity field. Within the single excitation subspace of this basis are two different kinds of states. 

The first kind of basis state represents the following configuration: the electromagnetic field mode is in its first excited state, while the molecular system is in the ground state, denoted as $|\mathrm{Ph}\rangle$. The second type of basis state describes the opposite configuration, where the excitation occurs on a single localized monomer fragment of the molecular system and the electromagnetic field mode is in the ground state, denoted as $|\mathrm{LE}_I\rangle$.


%
%

 The eigenstates of the TC Hamiltonian in the single excitation subspaces have an especially simple form if there is no detuning between the resonance frequency of the electromagnetic field mode and the excitation energie of the two-level systems ($\omega = \omega_{\mathrm{ph}} = \omega_{\mathrm{mol}}$) and are given by

\begin{equation}
    |\pm\rangle = \frac{1}{\sqrt{2}}\left(|\mathrm{Ph}\rangle + \frac{1}{\sqrt{N}}\sum_I^N |\mathrm{LE}_I\rangle\right)\label{eq:splitting},
\end{equation}
with eigenenergies $E_\pm = \hbar\omega \pm \hbar g \sqrt{N}$. These two states are the so-called upper and lower polaritons. Furthermore, there are $N-1$ degenerate eigenstates 

\begin{equation}
    |\mathrm{DS}_k\rangle = \sum_I^N c_{Ik} |\mathrm{LE}_I\rangle,\label{eq:ds}
\end{equation}
with eigenenergy $E=\hbar\omega$, where the coefficients have to fulfill the condition that $\sum_I^Nc_{Ik}=0$. Since these states do not have any contribution from the photonic state, they are often called dark states (DS). 


\subsection{Polaritonic interactions in FMO-LC-TDDFTB}
\begin{figure}[b!]
    \centering
    \includegraphics[width=\linewidth]{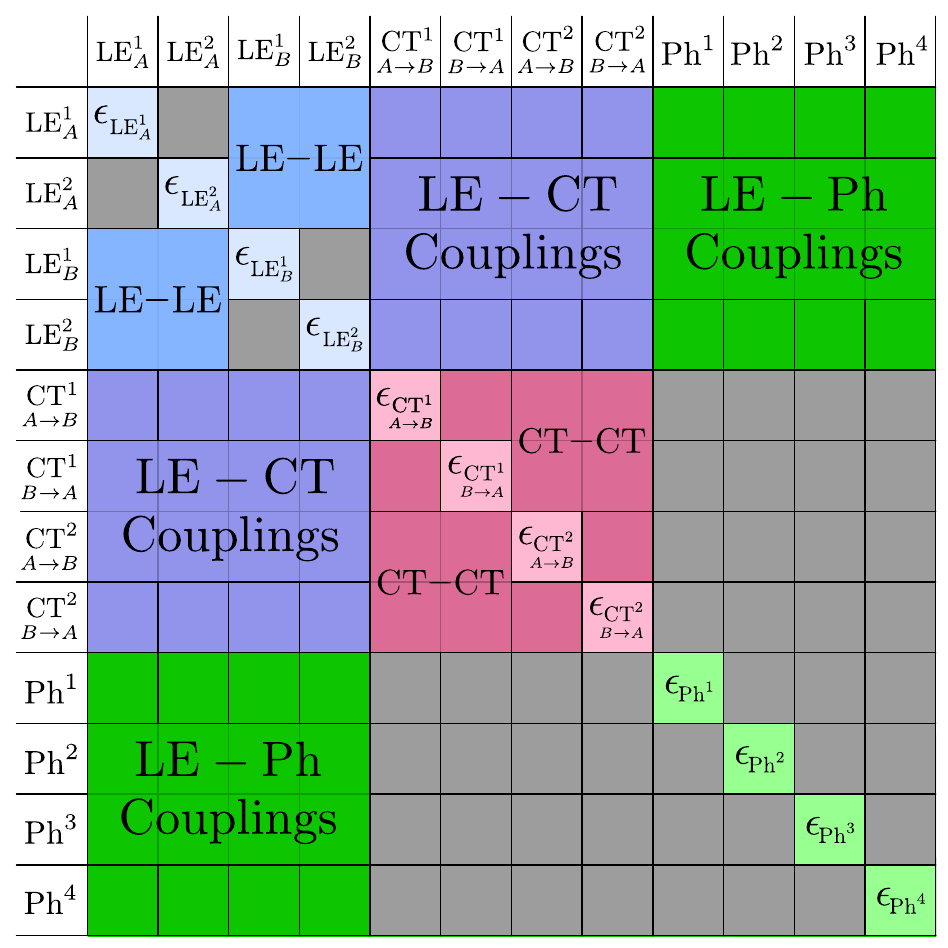}
    \caption{Depiction of the polaritonic Hamiltonian of a dimer system with two LE states for each fragment, two CT states and four photonic states. The gray matrix elements denote zero values.}
    \label{fig:exc_tc_hamiltonian}
\end{figure}
The generalized Tavis-Cummings model can be readily combined with the FMO-LC-TDDFTB formalism to account for strong-coupling between many electromagnetic field modes and many multi-state molecular systems. To this end, the quasi-diabatic basis of LE and CT states, which was introduced in section \ref{sec:FMO-LC-TDDFTB}, has to be expanded with the eigenstates of the electromagnetic fields. 

Thus, compared to the excited-state wavefunction in the FMO-LC-TDDFTB methodology (cf. Eq. \ref{eq:exc_state_wf}), the wavefunction in the polaritonic framework consists of locally excited, charge-transfer and photonic  basis states
\begin{equation}
\begin{split}
    \left|\Psi\right\rangle=&\sum_{I}^N \sum_{m}^{N_{\mathrm{LE}}} c_I^m \left|\mathrm{LE}_I^m \right\rangle+ \sum_{I}^{N} \sum_{J \neq I}^{N} \sum_{m}^{N_{\mathrm{CT}}} c_{I\rightarrow J}^m \left| \mathrm{CT}_{I \to J}^{m}\right\rangle \\
    &+ \sum_{\lambda} c_{\lambda} \left| \mathrm{Ph}_{\lambda} \right\rangle.
\end{split}
\end{equation}
In analogy to sections \ref{sec:Tavis-Cummings} and \ref{sec:FMO-LC-TDDFTB}, the first excited photonic state in the $\lambda$-th mode of the electric field is denoted by $\left| \mathrm{Ph}_{\lambda} \right\rangle$.


In order to calculate the excited states of the polaritonic system, the couplings between the LE, CT and photonic basis states are required. As indicated by Eq. (\ref{coupling}), the light-matter coupling of a molecular transition vanishes if the corresponding transition dipole moment is zero. Thus, CT states of different fragments generally do not couple to electromagnetic fields, since they do not possess a significant transition dipole moment
\begin{equation}
    \langle \mathrm{CT}_{I\rightarrow J}^m | \mathrm{H} \left| \mathrm{Ph}_{\lambda}\right\rangle \approx 0.
\end{equation}

Therefore, to a very good approximation, we only have to consider the coupling between LE states and electromagnetic field modes, which can be expressed by utilizing Eq. (\ref{coupling})

\begin{equation}\label{eq:LE-Ph-coupling}
    \langle \mathrm{LE}_I^m\ | \mathrm{H} \left| \mathrm{Ph}_{\lambda}\right\rangle = \hbar \sqrt{\frac{\hbar\omega_{c,\lambda}}{2\epsilon_0 V}} \Vec{\mu}_{eg}^{m(I)}\cdot\Vec{\epsilon}_\lambda,
\end{equation}
where $\Vec{\mu}_{eg}^{m(I)}$ is the transition dipole of the \textit{m}-th locally excited state of monomer \textit{I}.
The total Hamiltonian of the strong-light matter coupling in the framework of the FMO-LC-TDDFTB method is shown in Fig. \ref{fig:exc_tc_hamiltonian} for an example dimer system. 

To obtain the excited states of the coupled system between the cavity and the molecular aggregate, the total Hamiltonian is diagonalized. Subsequently, the intensity of the excited polaritonic states are calculated according to the following expression
\begin{equation}
    f_i = \sum_{\lambda} \omega_i |c_{\lambda}|^2,\label{eq:intensity}
\end{equation}
where $\omega_i$ is the energy of the respective state and $c_{\lambda}$ is the coefficient of the photonic basis state of the $\lambda$-th mode.
\section{Computational details}\label{sec:comp_details}
The computational details of all calculations that were performed in this work are presented in this section.

The DFTB parameter set ob2\cite{vuong_parametrization_2018} was utilized in all calculations of this work. A long-range radius of $3.03$~$ \mathrm{a_0}$ was employed in the framework of LC-DFTB. 

We used the Mercury program\cite{macrae_mercury_2020} to generate the molecular structure of 125 tetracene monomers from the crystal structure of tetracene \cite{holmes_nature_1999}.

In case of the FMO-LC-TDDFTB calculations, the number of locally excited states was set to four, whereas the number of charge-transfer states was set to one. 

For the calculations of the polariton dispersion of the excited-states of the tetracene assembly, we performed a scan of the photon energy for the three polarizations of the electric field along the x-, y-, and z-axis. We employed one photon mode to simulate the strong light-matter coupling between the cavity photon and the first two excited states of the tetracene molecules. 
\begin{figure}[t!]
    \centering
    \includegraphics[scale=0.85]{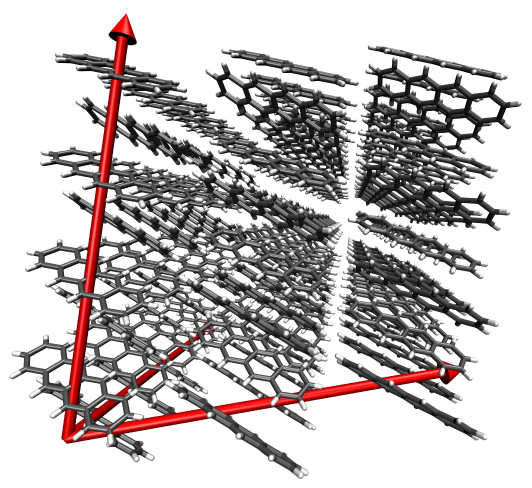}
    \caption{Depiction of the molecular structure of a part of the tetracene crystal structure with the crystallographic axes.}
    \label{fig:tetracene_cluster}
\end{figure}
\section{Results}\label{sec:results}
In this section, the influence of the polarization vector of the electric field on the polaritonic excited states will be investigated by scanning the angle of the polarization of the electric field of a cavity mode that is coupled to an aggregate of 125 tetracene molecules, which is shown in Fig. \ref{fig:tetracene_cluster}, in the xy-plane. Furthermore, polaritonic spectra are calculated for three different tetracene aggregates to study the influence of the system size on the excited-state energies of the upper and lower polaritonic branches. Subsequently, the polaritonic dispersion of the tetracene system that is composed of 125 monomers is examined by varying the photon energy of the cavity modes. This scan is performed for the x-,y- and z-polarization of the electric field of the cavity photons.
\subsection{Angle dependence of the polaritonic spectrum}
\begin{figure}[b!]
    \centering
    \includegraphics[width=\linewidth]{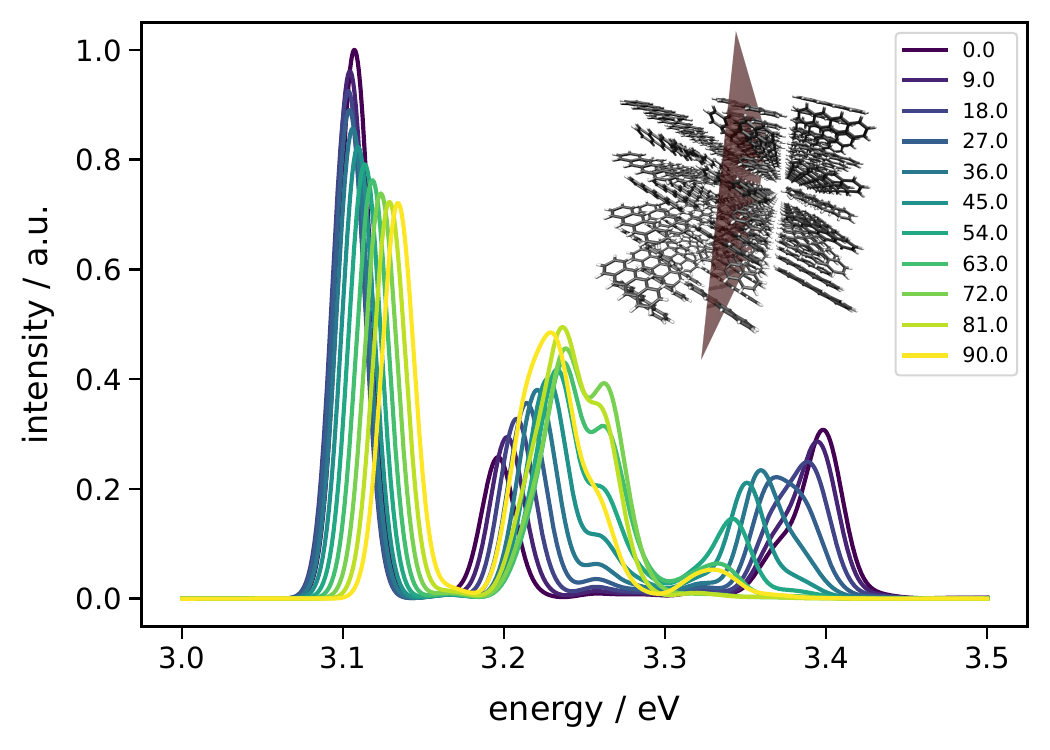}
    \caption{Scan of the polarization angle of the electric field of the cavity mode. The inset picture of the tetracene cluster shows the xy-plane.}
    \label{fig:rotation_scan}
\end{figure}
To explore the influence of the polarization vector of the electric field of the cavity photons on the excited polaritonic states of the tetracene-cavity system, the polarization angle of the electric field is scanned in the xy-plane. To this end, the energy of the cavity photon is set to $3.2$~eV, which approximately corresponds to the excited-state energy of the first excited state of the tetracene monomers. The polarization vector at $0^\circ$ ($90^\circ$) corresponds to pure x-polarization (y-polarization).
The resulting spectra of the polaritonic excited-states of the tetracene-cavity system for various polarization angles are depicted in Fig. \ref{fig:rotation_scan}.
At all angles the polaritonic spectra exhibit three main absorption peaks, which are centered around ca. $3.1$, $3.2$, and $3.4$~eV at $0^\circ$. In the following, we will refer to underlying excited polaritonic states as lower, middle, and upper polariton. The appearance of three energetically well separated peaks in the absorption spectra show that there are at least two different electronically excited states of the tetracene crystal involved in the formation of the polaritonic states. Actually, the small shoulders at some of the three main absorption peaks hint that there are more states included in the formation.

If now the polarization vector is rotated from the pure x-direction to the y-direction, the energy of the lower and middle polariton shift towards higher energies. Conversely, the energy of the upper polariton is shifted to lower energies. Overall, the energetically decreasing gap between the lower and upper polaritonic states directly shows that the coupling strength between the cavity mode anf the excited states of the tetracene aggreagte is strongest for purely x-polarized electomagnetic field modes. 

In addition, an increase in the polarization angle leads to a decline of the intensity of the upper and lower polaritonic states, which is caused by the decreasing contributions of the cavity states. However, in conjunction of the declining intensities of the polaritonic states at 3.1 and 3.4~eV, the polaritonic state between 3.2 and 3.3~eV shows an increase in the oscillator strength.
The scan of the polarization angle clearly shows that the highest coupling between the excited states of the tetracene molecules and the photon modes of the cavity is reached for the linear polarization of the electric field along the x-axis.
\subsection{Influence of the system size on polaritonic spectra}
\begin{figure}[t!]
    \centering
    \includegraphics[width=\linewidth]{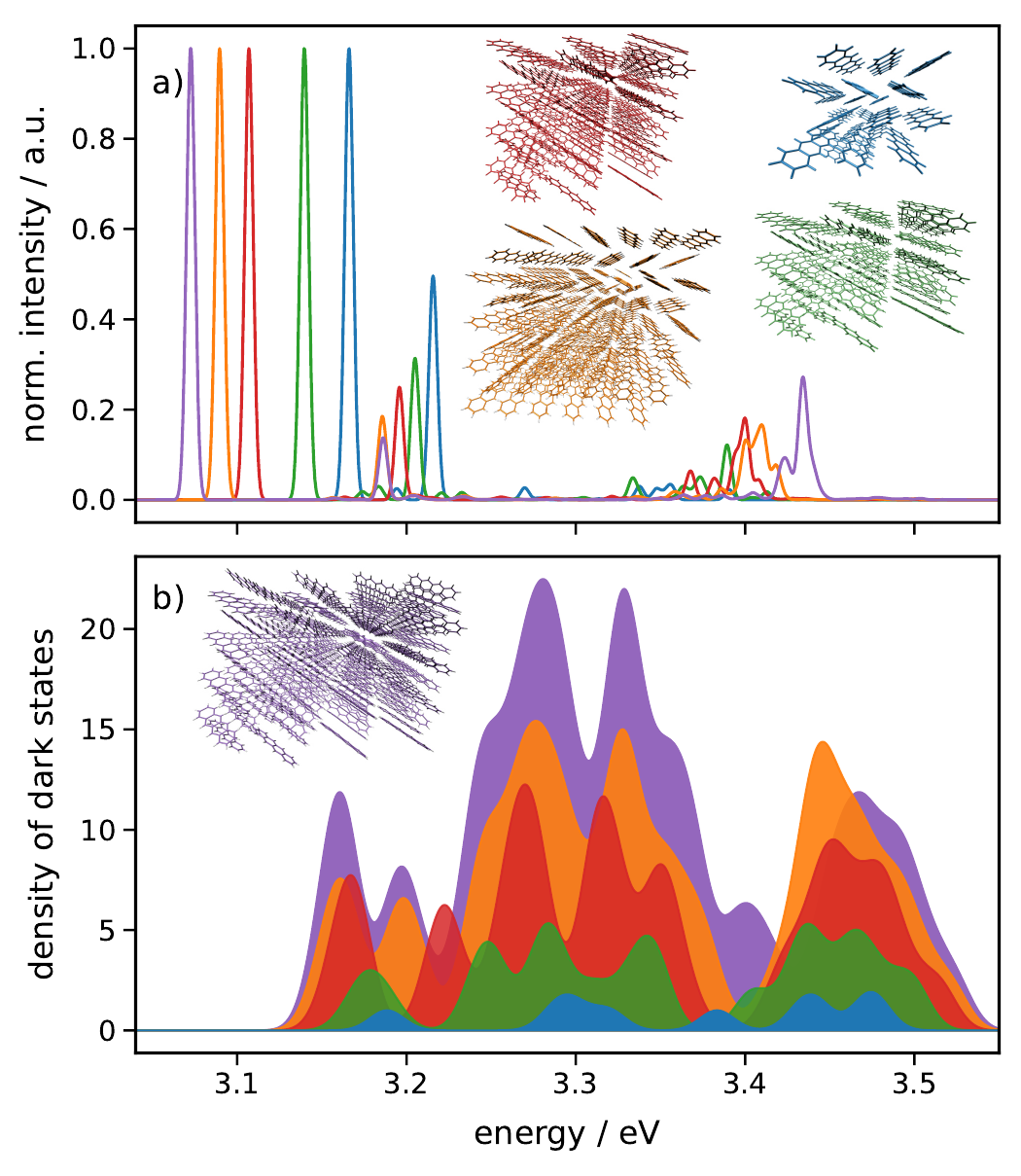}
    \caption{a) Polaritonic spectra of five different tetracene aggregates. b) density of states of the dark states of the tetracene aggregates.}
    \label{fig:splitting_comparison}
\end{figure}
In order to explore the influence of the system size on the excited state energies of the polaritonic system, the spectra of four different tetracene aggregates, which contain 22, 65, 125, 167 and 224 monomers, are compared. The electric field of the cavity mode that interacts with the electronically excited states of the tetracene system is polarized in the x-direction. The energy of the photon mode is set to ca. 3.2~eV.

The polaritonic spectra of the five tetracene aggregates are depicted in Fig. \ref{fig:splitting_comparison}a. As expected for larger systems, the spectra show an increasing red shift of the excited-state energies in conjunction with the increase in system size. This is caused by the J-coupling\cite{jelley_spectral_1936,jelley_molecular_1937,scheibe_uber_1937,tuszynski_mechanisms_1999,walczak_exchange_2008} between the $S_1$ states of the monomer fragments, which is enhanced by the increase in monomer fragments. 

However, aside from the red-shift of the polaritonic spectra, an increase of the splitting between the upper and lower polaritonic branches can be observed. This phenomenon is expected, because the splitting between the upper and lower polaritons scales with the factor $g \cdot \sqrt{N}$ (cf. Eq. \ref{eq:splitting}). 

In addition, the density of states of the dark states of the tetracene fragments is depicted in Fig. \ref{fig:splitting_comparison}b. In general, there are two different kinds of dark states, which can arise in polaritonic systems. The first kind of dark states, usually referred to as the uncoupled molecules, are molecular excited states, which do not couple to the electromagnetic fields. Usually these do not couple to the field, because either they have no transition dipole moment at all or their transition dipole moment is oriented perpendicular to the polarization vector. The second kind of dark states arise as eigenstates of the Tavis-Cummings Hamiltonian (\textit{cf.} Eq. \ref{eq:ds}). In this case, the quasi-diabatic states, which contribute to the eigenstate do actually couple to the electromagnetic field. However, they are linear combinations of these states such that they do not have any contribution from the photonic states. Thus, according to Eq. \ref{eq:intensity} they are not visible in an absorption spectrum. While the dark states are always energetically located between lower and upper polariton, uncoupled molecular states can arise at various energies. From Fig. \ref{fig:splitting_comparison}b it is evident that the energetically lowest band of dark states shifts with growing aggregate size towards lower energies such that it stays in between the middle and lower polariton. We conclude that these dark states have to emerge within the strong coupling process. Contrary, the band of dark states at ca. 3.45~eV can only be uncoupled molecular states, since they lie beyond the polaritonic states. However, the major part of the density of dark states lies between those two extremes located around 3.3~eV. For these states one can not unambiguously decide, if they are uncoupled molecular states or if they arise due to the strong-coupling. Certainly, it is a combination of both.

\begin{figure}
    \centering
    \includegraphics[width=\linewidth]{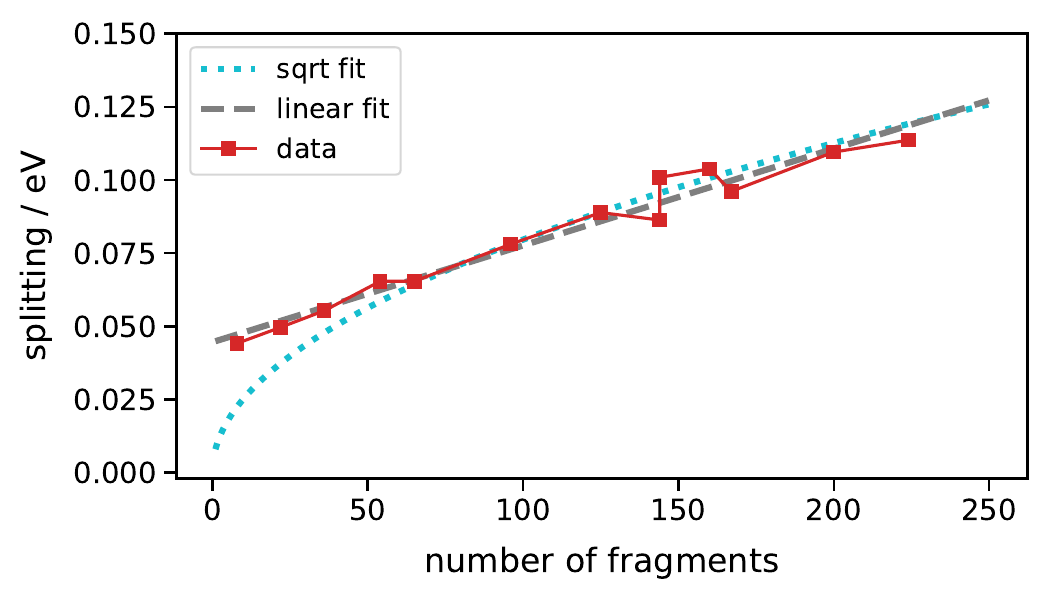}
    \caption{Influence of the number of fragments on the splitting of the polaritonic states.}
    \label{fig:splitting}
\end{figure}

To investigate the influence of the size of the tetracene aggregate on the splitting between the polaritonic states, the splitting between the first two polaritonic states is compared and depicted in Fig. \ref{fig:splitting}. It can be observed that different aggregates of the crystal structure, which consist of the same number of fragments, show different values for the splitting between the polaritonic branches. This effect is caused by the disparity in the transition dipole moments of the excitonic states between the various structures, which leads to smaller or greater coupling to the cavity mode.
As stated previously, the splitting of the polaritonic states is influenced by the J-coupling between the tetracene fragments, which is enhanced in conjunction with the increase in the number of monomers. As is commonly known, it also scales with the factor $g \cdot \sqrt{N}$, thus a combination between both phenomena is expected. The comparison of the splitting for thirteen tetracene aggregates indicates an approximately linear increase of the splitting with growing numbers of tetracene fragments, and thus, the observed scaling is different from $\sqrt{N}$, which validates the assumption of a combined scaling behaviour. However, the energy of the photon mode is approximately set to the $S_1$ energy of the monomer fragments. If the energy of the photon is adjusted to the resonance frequency of each tetracene aggregate, a different scaling behaviour might occur.

\subsection{Polariton dispersion of tetracene aggregates}
To investigate the dependence of the polariton dispersion of tetracene aggregates in respect to the polarization of the electric field modes of the cavity, a scan of the photon energy of the cavity mode is performed.
\begin{table}[t!]
\caption{Absolute values of the transition dipole moment of a tetracene monomer (in atomic units).}
\label{tab:transition_dipole_moments}
\begin{tabular}{c|ccc}
\toprule
state & x-axis & y-axis & z-axis  \\ \midrule
$S_1$  & 1.01  & 0.52  & 0.06  \\
$S_2$  & 0.24  & 0.28  & 0.95  \\ \bottomrule
\end{tabular}
\end{table}
To this end, the photon energy is scanned between the values of 2.6~eV and 4.8~eV in order to obtain the polaritonic states, which correspond to the interactions between the cavity mode and the first two electronically excited states of the tetracene fragments. The scan is performed for the x-,y- and z-polarization of the electric field of the photon modes. 
\begin{figure}[b!]
    \centering
    \includegraphics[width=\linewidth]{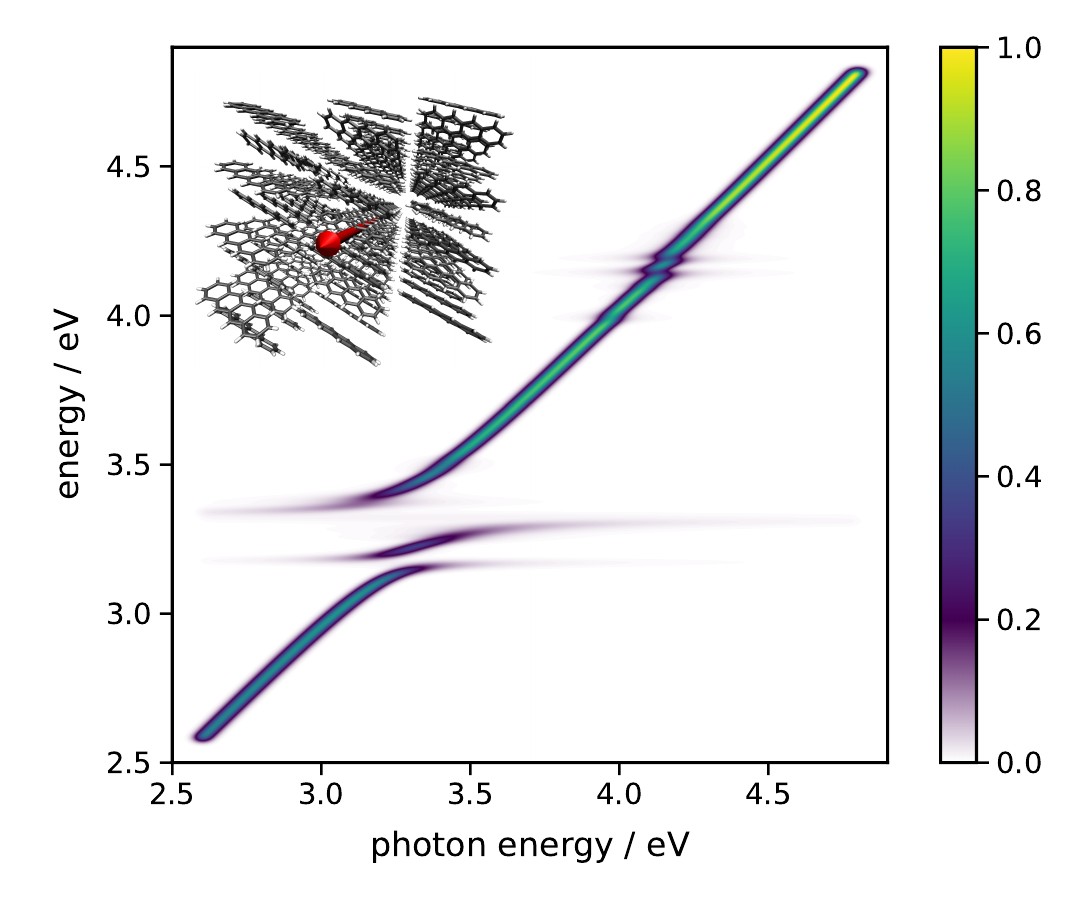}
    \caption{Polariton dispersion of the tetracene assembly for a polarization of the electric field along the x-axis. The polarization vector of the electric field is depicted together with the tetracene cluster.}
    \label{fig:x_dispersion}
\end{figure}

Due to the fact that the coupling strength of the interaction between the cavity modes and LE states is dependent on the length and direction of the transition dipole moment vector in the direction of the polarization vector (cf. Eq. (\ref{eq:LE-Ph-coupling})), the polariton dispersion differs for each of the three cartesian polarization directions of the electric field modes. Therefore, the values of the transition dipole moment of the tetracene monomer, which are shown in Tab. \ref{tab:transition_dipole_moments} for the first two electronically excited states, are an useful estimate for the coupling strength of light-matter interactions, and thus, the splitting of the upper and lower polaritonic states.
The polariton dispersion of the tetracene aggregate for an electric field vector that is polarized in the x-direction is shown in Fig. \ref{fig:x_dispersion}. As the tetracene monomers have the highest transition dipole moment value in x-direction for the first electronically excited state, the largest splitting between the upper and lower polaritonic branches is observed at a photon energy of ca. 3.2 eV, which corresponds to the transition to the $S_1$ state. However, due to the relatively low value of the transition dipole moment along the x-direction for the $S_2$ state, the coupling between the cavity mode and the tetracene aggregate is small. 
\begin{figure}[t!]
    \centering
    \includegraphics[width=\linewidth]{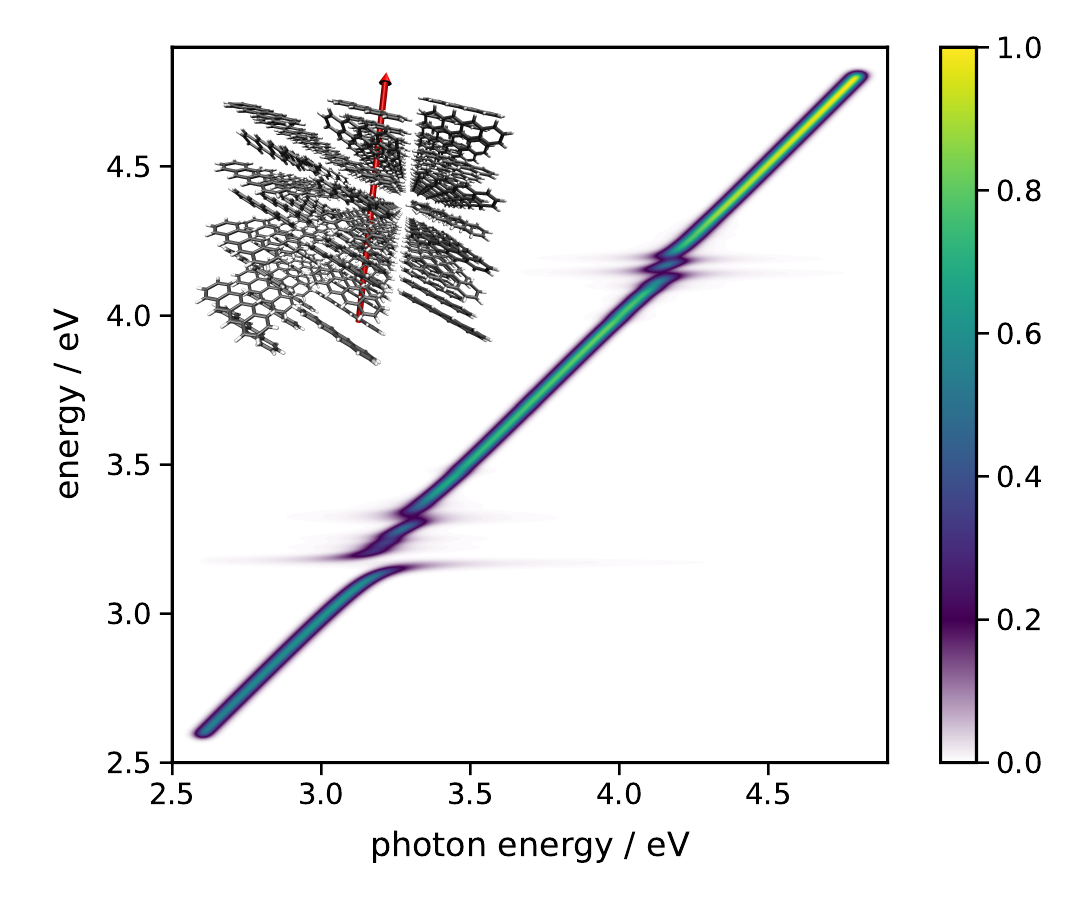}
    \caption{Polariton dispersion of the tetracene assembly for a polarization of the electric field along the y-axis. The polarization vector of the electric field is depicted together with the tetracene cluster.}
    \label{fig:y-dispersion}
\end{figure}
\begin{figure}[b!]
    \centering
    \includegraphics[width=\linewidth]{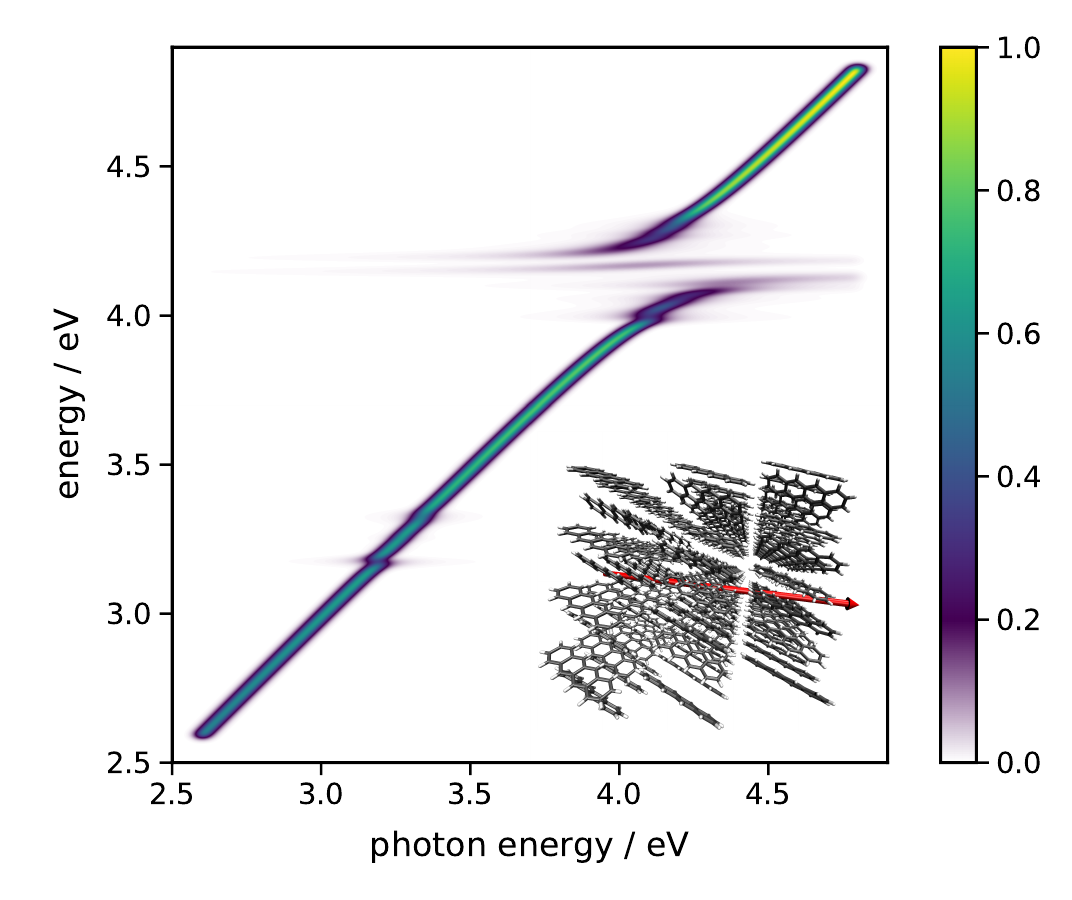}
    \caption{Polariton dispersion of the tetracene assembly for a polarization of the electric field along the z-axis. The polarization vector of the electric field is depicted together with the tetracene cluster.}
    \label{fig:z-dispersion}
\end{figure}

The polariton dispersion for the electric field polarization along the y-axis shows similar results. The transition dipole moment for the transition to the $S_1$ state has a higher value than the transition to the $S_2$ state (cf. Tab. \ref{tab:transition_dipole_moments}). Subsequently, the splitting between the upper and lower polaritonic branches that correspond to the excitation to the $S_1$ states of the tetracene fragments, which is shown in Fig. \ref{fig:y-dispersion}, is larger than the splitting for the $S_2$ states.

Compared to the polaritonic couplings for the electric field polarizations along the x- and y-axes, the polarization along the z-axis yields the opposite results. As shown in Tab. \ref{tab:transition_dipole_moments}, while the transition dipole moment along the z-axis shows the lowest values for the excitation to the $S_1$, the value for the transition to the $S_2$ is the highest. Subsequently, the splitting of the upper and lower polaritonic branches is really small for the $S_1$ states and shows a large value for a photon energy of approximately 4.1 -- 4.2 eV, which corresponds to the $S_2$ states of the tetracene aggregate (cf. Fig. \ref{fig:z-dispersion}).

\section{Conclusions and outlook}\label{sec:conclusions}
In this paper, we have presented a novel methodology to calculate the polaritonic excited-state spectra of large molecular assemblies that consist of hundreds of monomer fragments by combining the excitonic Hamiltonian of our FMO-LC-TDDFTB method with a generalized TC Hamiltonian. We extended our quasi-diabatic basis of LE and CT states with cavity photon modes to construct a polaritonic Hamiltonian, which incorporates the excitonic interactions between all molecular fragments. We implemented our new methodology in our software package DIALECT, which is available on Github.\cite{noauthor_dialect_2024}

We demonstrated the capability of our method in simulating the dependence of the polaritonic spectra on the polarization angle of the electric field vector of the cavity photons by calculating the spectra of a cavity coupled tetracene systems of 125 monomer fragments. Additionaly, we investigated the influence of the system size of tetracene aggregates on the excited-state energies of the upper and lower polaritonic branches, which showed an increased energy gap and a red shift of the excited-state energies in conjunction with the increase in aggregate size. 

We also demonstrated the potential of our methodology in simulating the plexciton dispersion of large molecular aggregates that are coupled to cavity photon modes. To this end, we calculated the polaritonic excited-state spectra for a tetracene cluster of 125 monomer fragments and scanned the energy of the photon mode to obtain a two-dimensional representation of the dispersion of the polaritonic branches.

In the future, we plan to extend our recent work on the quantum-classical nonadiabatic excitonic Ehrenfest dynamics in large molecular systems\cite{einsele_nonadiabatic_2024} to simulations of dynamical processes induced by the coupling of molecular systems to microcavities.
For this purpose, we will implement the analytical gradients of the couplings between the photonic states of the cavity and the LE states of the molecular fragments. 

\section*{Conflicts of interest}
\vspace*{-2ex}
There are no conflicts to declare.
\section*{Acknowledgments}
\vspace*{-2ex}
We gratefully acknowledge financial support by the Deutsche Forschungsgemeinschaft via the grants MI1236/6-1 and MI1236/7-1. L. N. P. acknowledges the support of the Kekul\'e stipendium of the FCI.
\section*{Data availability}
\vspace*{-2ex}
The data of the calculations that were performed in this work is available at github \url{https://github.com/mitric-lab/FMO_LC_TDDFTB_strong_light_matter_coupling}.
\section*{References}
\vspace*{-2ex}

\bibliography{references}

\end{document}